# Global picture of OAI-PMH repositories through the analysis of 6 key open archive meta-catalogs

Authors: Arnaud Gaudinat, Jonas Beausire, Megan Fuss, Elisa Banfi, Julien Gobeill, Patrick Ruch

## Abstract

By the end of the late 90's the Open Archives Initiative needed direction to insure its improvement and thus, created the Open Archives Initiative Protocol for Metadata Harvesting (OAI-PMH) standard. The movement showed a rise in popularity, followed by a decline then a relative stabilization. This process was essentially a way to ensure the viability of open archive repositories. However, a meta-catalog containing an ensemble of repositories was never established, which lead to confusion of what could be found in said catalogs.

This study ultimately aims to find out what repository content can be found and where with the use of the 6 key meta-catalogs. Although they undoubtedly have numerous limitations pertaining to the available data, this article seeks to compare the common data in each meta-catalog and estimates which repositories are found within them (with approx. less than 1% in common within the 6 meta-catalog).

Decisively, this paper identifies the need to collate this data (with a total of 42.3% OAI-PMH repositories specific to each meta-catalog) and improve current search tools, hence portraying the benefits of a comprehensible single unifying meta-catalog for end users.

## Introduction

The Santa Fe Convention (Sompel 2000) of October 1999 brought together open archives managers around the Open Archives Initiative (OAI). It was the starting point and the foundation of a technical and organizational framework for a standard that improves the interoperability and the exchange of data from archives. This standard officially became the Open Archives Initiative Protocol for Metadata Harvesting (OAI-PMH) in July 2001 (Suleman 2001). This new protocol (currently v 2.0) offers repositories around the world a common computer base for collecting metadata, hence promoting the exchange, meshing and interoperability of data and archives (Open Archives 2001).

At the dawn of the twenty-first century, the scholarly authors that gathered at the Santa Fe Convention were not mistaken when they foresaw an important expansion of open archives. Today, thousands of repositories (data providers) have emerged all over the globe, whether they are institutional, governmental, or scientific repositories (Norris, Oppenheim, Rowland 2008), they are for the most part multidisciplinary, contain predominantly English sources and are very heterogeneous (Pinfield et al. 2014). Among them, a constellation of meta-repositories (referred to as meta-catalogs in this study) that propose to refer and organize institutional repositories (data providers), listed via a single platform. These efforts of centralization help prevent users of these archives from encountering a dispersion and offer visibility to content too often scattered or isolated: "[…] there is certainly an argument that OA repositories containing research outputs can deliver a relative advantage to organizations by making the outputs of their staff more visible and therefore increasing academic and societal impact." (Jones, Andrew, MacColl 2006).
Today however, the increase of these centralization initiatives demonstrates a new landscape of meta-catalogs, each of which seeks to focus and collate relevant repositories. This important development also applies to the institutional context of the Open Access movement (OA) which structures the policies of these meta-catalogs (Xia 2012), therefore also changing academic practices.

In this context, questions remain concerning the updates of these meta-catalogs, the possible redundancies in the deposits listed and the disorder that they would imply.

Faced with these two observations, that of an expansion of repositories and a possible disorder of their contents, this article proposes to establish an inventory of six important meta-catalogs: OAIster (Hagedorn 2003), OpenArchives, OpenDOAR, Registry of Open Access Repositories (ROAR), OpenAIRE (Manghi 2010) and The University of Illinois OAI-PMH Data Provider Registry (Illinois). Several questions arise: what can be found there? What data is truly accessible? What is their precise coverage? Is there a tool to access this repository ecosystem? If not, could it be built?

Thus, two hypotheses can be formed:
1) The meta-catalog content should be relatively similar between the up to date meta-catalogs (ROAR, OAIster, OpenDOAR, and OpenAIRE for Europe).
2) Most of the old meta-catalog content should be included.

In order to answer these questions, this study will be divided into two main parts and will either confirm or disprove our hypotheses. The first part will globally evaluate the vitality of OAI-PMH through bibliometric and webometric approaches, and also by directly using descriptive data available on Internet. The second part will analyze the content of six key meta-catalogs with a descriptive method and an exploratory method based on systematic and automatic approaches. This will reveal the fundamental differences and salient features of the studied meta-catalogs. Therefore, the results will allow an accurate reflection of the state of OAI-PMH and its usage today.

The Open Archives Initiative is frequently associated with the open access publishing movement and they are indeed related through the means of access and sharing. There are numerous evaluations of the state of open access around the world (Björk 2010 and 2014, Morisson 2012), they mainly address the differences between open access and non-open access through numerous characteristics such as impact, discipline, cost, evolution and geographical specificity. However, this study is not an evaluation of the situation of open access but more of an evaluation of the state of open archives. On this subject, Schöpfel conducted an interesting study in 2008 on the situation relating to grey literature in France. Nonetheless, the main subject studied here pertains to the situation of one of the most popular interoperability standards for sharing catalogs around the world, the OAI-PMH protocol. To our knowledge, it has not yet been studied to this extent and this paper presents an original overview of OAI-PMH's current usage and some of its related challenges.

**Definition and perimeters of this study**

This article studies the *"catalog of catalogs"* which could be a confusing term. The upper level, or *"catalog of catalogs"*, will be called meta-catalog to avoid ambiguities. Each studied meta-catalog contains a list of open archive repositories at International level or European level. For the purpose of this study, only repositories containing records available through the OAI-PMH protocol are utilized. These repositories will be referred to as OAI-PMH repositories (like in Pinfield 2014) or merely as repositories for the rest of this document (OAI-PMH repositories could be called OAI-PMH servers or archives in other articles).

## Methodology

**Part 1 - Vitality of OAI-PMH**

**Bibliometrics and webometrics approach**

The first step is evaluating the activity of said topic through time. For this, classical bibliometric reviews as well as webometric reviews were used. For the bibliometric side, Google Scholar was queried for specific years that only contained the 'oai-pmh' keyword in the title, the corresponding number of article per year was then recorded. It can be noted that Google Scholar provides good bibliometric indicators (Franceschet 2010), meaning fifteen queries from 2001 to 2016 (results in figure 2). Regarding the webometric part, Google Trends was queried at a global level mainly using the 'oai-pmh' keyword but also two additional open archive related topics characterized by the keyword 'oai-ore' and the keyword 'ResourceSync'. Google Trends[1] is based on the number of queries previously made by Google Web search engine users. This tool is thoroughly compelling because it represents an interest of ongoing legitimate topics for a given period. It is also a relevant tool in our study with cases such as non-polysemy keywords and equivalent number of words in the queries.

**History of meta-catalog evolution[2]**
Another way to assess the vitality of the OAI-PMH world is to evaluate the increasing number of OAI-PMH repositories through time. Unfortunately, only OpenDOAR provides an interface to obtain this information. It was reformatted in figure 3 by manually collecting the number of repositories for each year in January (values are rounded up by the interface to 3 digits). However, the Wayback Machine (from Internet Archive) was used to manually track the number of repositories as they were archived and then indicated on the main page of the Open Archive Initiative Web site (figure 4). By using this method, the figures from 2004 to 2016 could be obtained (due to the first capture available on Internet Archive the selection of each year was made in October). The possibility of using the Wayback Machine method for the four others meta-catalogs was evaluated but the first available capture dates were too recent to be useful. ROAR has a built-in tool to help obtain the number of repositories per year, although the data is not of historical nature but that of current information.

**Part 2 – Evaluation of key open archive meta-catalogs**

**Establishing a list of the key open archive meta-catalogs**

---

[1] https://www.google.com/trends/
[2] http://roar.eprints.org/view/year/).

The list of studied meta-catalogs was obtain by means of literature reviews and empirical searches fulfilled using Google. One of the requirements was to have the most exhaustive meta-catalogs available on a global scale (except for OpenAIRE it was nonetheless deemed relevant to compare it to the other meta-catalogs). Another way to check the availability of the other meta-catalogs, was to query Google using some of the known OAI-PMH repository URLs (pmc, rero, hal and others). Subsequently, this list of meta-catalogs is not exhaustive but certainly a suitable representation on a global level. Incidentally, the use of six meta-catalogs provides a viable sample size for this study, particularly for building a comprehensible Venn diagram.

In order to establish a comparative table (table 2) some administrators of the selected meta-catalogs (OAIster, OpenArchives) were directly contacted. This data was sought out to complete the fractional information obtained on the websites of the meta-catalogs or in grey literature.

**Acquiring a list of OAI-PMH repositories**

The procurement of the full list of OAI-PMH repositories for each meta-catalog can be considered the heart of this study. It needed to be obtained in order to evaluate the specifics or commonalities between them. OAI-PMH being an operational system, this study is based on the comparison of URLs of each repository (by way of identifiers) which are for instance less ambiguous than the names of the repositories. This acquisition of this list was done in the month of January 2017.

Unfortunately, there are no uniform formats or services that can be employed to obtain the list of repositories for each meta-catalog. OpenDOAR, ROAR, OpenArchives and Illinois each have their own xml web services that can be used to query their content (Table 1). For the four meta-catalogs in question, the URLs of the OAI-PMH servers were extracted from the XML services by applying regular expression rules. Even though OAIster has several web services mainly relating to WorldCat, there was no obvious service to obtain *its* list of repositories. This list was obtained by scraping OAIster's html page with phantomJS's Web client which can interpret Javascript on the fly. OpenAIRE meta-catalog has several dedicated, advanced and standardized web services (such as the meta-catalog itself, a OAI-PMH interface). However, obtaining the complete OAI-PMH repository URL listing was not obvious at all. It was then decided that a two-step scraping method should be used mfor OpenAIRE. Firstly, by acquiring the full list of data providers from the URL specified in HTML (table 1), then by selecting the 'Show All entries' and scraping the OpenAIRE URL of each the data provider items. Secondly, by accessing each of these URLs and applying a regular expression to extract the OAI-PMH repository URL from the HTML page of the data provider in question.

*Table 1: Scraping the list of OAI-PMH repositories*

|  | url | Simplified Regular Expression |
|---|---|---|
| OpenDOAR | http://www.opendoar.org/api13.php?all=y | <rOaiBaseUrl>(.*?)<\/rOaiBaseUrl> |
| ROAR | http://roar.eprints.org/rawlist.xml | <oai_pmh>(.*?)<\/oai_pmh> |
| OpenArchives | http://www.openarchives.org/pmh/registry/ListFriends | <baseURL(.*?)>(.*?)<\/baseURL> |
| Illinois | http://quest.library.illinois.edu/registry/ListAllAllRepos .asp?format=xml | <baseURL(.*?)>(.*?)<\/baseURL> |

| OAIster (OCLC) | http://www.oclc.org/oaister/contributors.en.html | \<strong\>OAI base:\<\/strong\>(.*?)\<\/p\> |
| OpenAIRE step 1 | https://www.openaire.eu/search/data-providers | \<a href=\"(\/search\/dataprovider)(.*?)\"\> |
| OpenAIRE step 2 | https://www.openaire.eu/search/dataproviders/* | \<a class=\"custom-external\" target=\"_blank\" href=\"(.*?)\"\>(.*?)\<\/a\> |

Some of the catalogs could possibly have several different URLs per item, if so, any additional URLs were considered as different OAI-PMH repositories.

**Simple normalization and deduplication of URLs for exact matching**

Some of the collected URLs have query string parameters which are inadequate for this study. This step removed the query string (if existent) of the URL (i.e the part after the question mark) which allowed the conception of an OAI-PMH query to check the status of the server as is described in the next section. During this step, exact duplicate URLs were also removed. In fact, the normalization and deduplication of URLs were minimal here in order to maximize the chance of reaching the OAI-PMH service without having to change the original URL excessively.

**Checking the OAI-PMH availability**

Then the status or availability of all the yielded URLS was checked through a standardized OAI-PMH query: "url-of-oai-pmh-repository" plus "?verb=Identify". The http status code was kept, and if available so was the repository name, the protocol version and the earliest date stamp.

**Stronger normalization and deduplication of URLs for comparison**

Finally, a stronger normalization and deduplication was applied to the URLs of the active OAI-PMH server to allow the comparison of URLs between different meta-catalog. For instance, the same repository could have a https protocol indication in one meta-catalog while having a http protocol indication in another meta-catalog. The URLs were then normalized and deduplicated to avoid similar identifiers for the same service. Then "http://", "https://", "www" and finally "/" were removed as done by McCown (2006).

# Results

**OAI-PMH vitality**

In figure 1, a rapid interest in OAI-PMH can be observed after its first publication in January 2001. With a total of 333 publications, there is the peak of 45 publications in 2006, thereafter, that number continues to decrease regularly until 2016 with 10 publications. Obviously, OAI-PMH is not currently an important active topic of research undoubtedly due to a combination of the fact of being well established and to a somewhat loss of interest.

Regarding literature on the topic of OAI-PMH, it is progressively less numerous than in the mid-2000s but it is also highlighted by the cross-cutting use of the protocol. It examines, for example, the users' engagement to the resources collected by the OAI-PMH Protocol (Allison 2016), ways to improve and simplify the harvesting process (Ong, Leggett, 2005) and even the prospect of a single collection of global data providers (Goebert, Harriehausen-Mühlbauer, Furnell 2014).

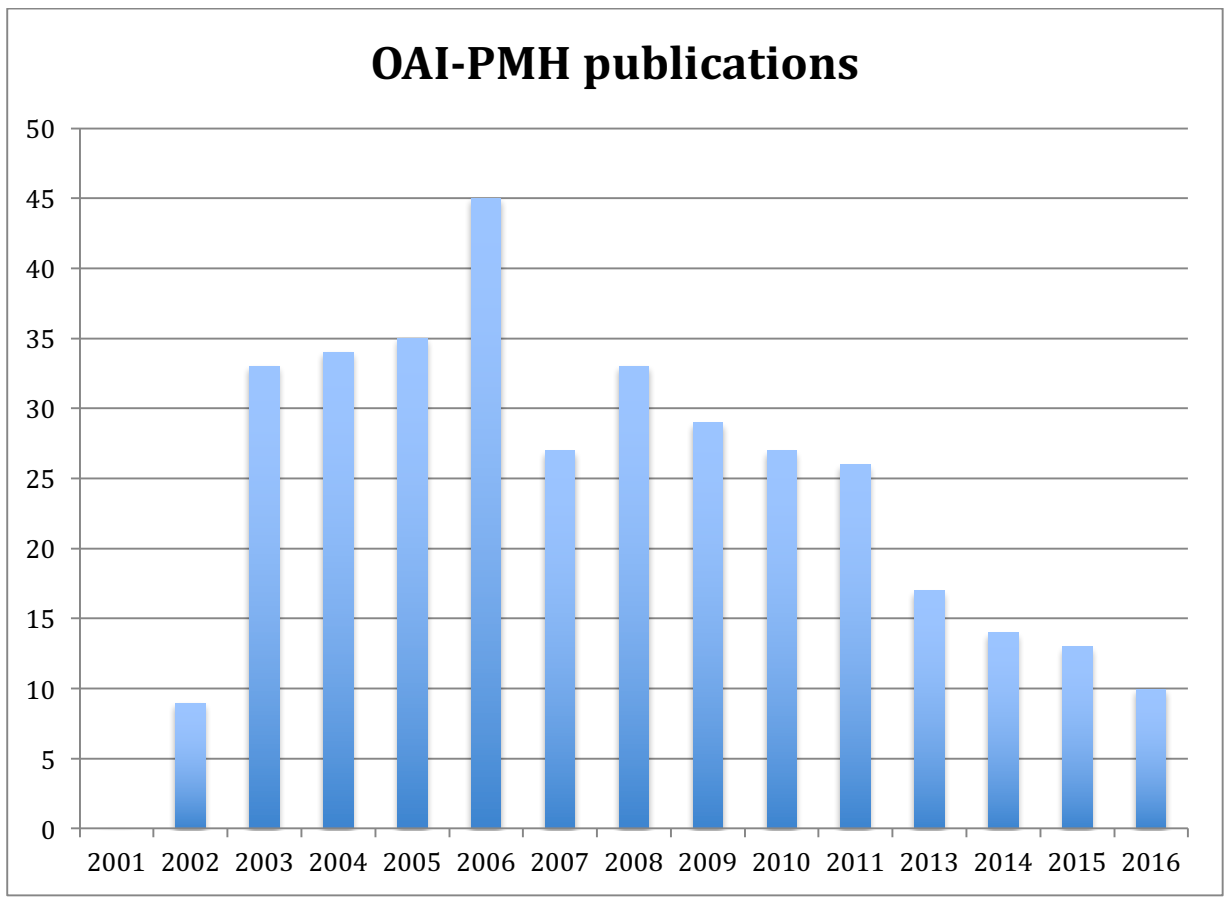

*Figure 1: "OAI-PMH" publication in title from Google Scholar (January 2017)*

Figure 2 confirms this trend with a continuous decrease of the 'oai-pmh' query whilst using Google search engine once again from 2006 to 2016. However, more recent initiatives in the domain such as OAI-ORE and ResourceSync haven't received as much interest as OAI-PMH when using the Google Trends measuring method (figure1) and Google Scholar publication (with the same method, only 3 for both OAI-ORE and ResourceSync).

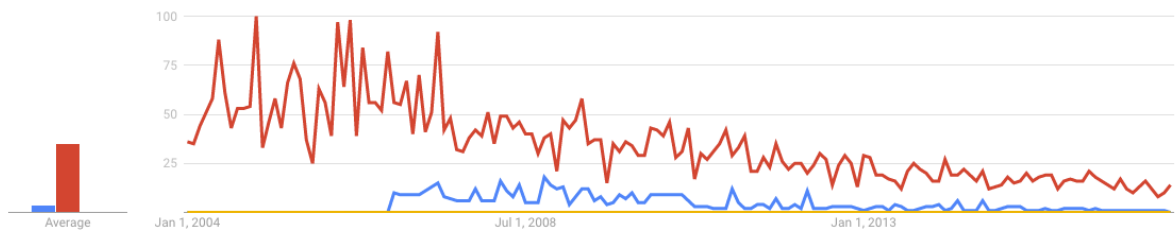

*Figure 2: OAI-PMH(red), OAI-ORE (blue) and ResourceSync (yellow) queries interest from Google Trends (January 2017)*

Figures 3 and 4 respectively indicate that there is a continuous growth of OAI-PMH repositories for OpenDOAR and OpenArchives from 2004 to 2016. Despite its old interface and age, the OpenArchives meta-catalog is surprisingly still very active when compared to OpenDOAR.

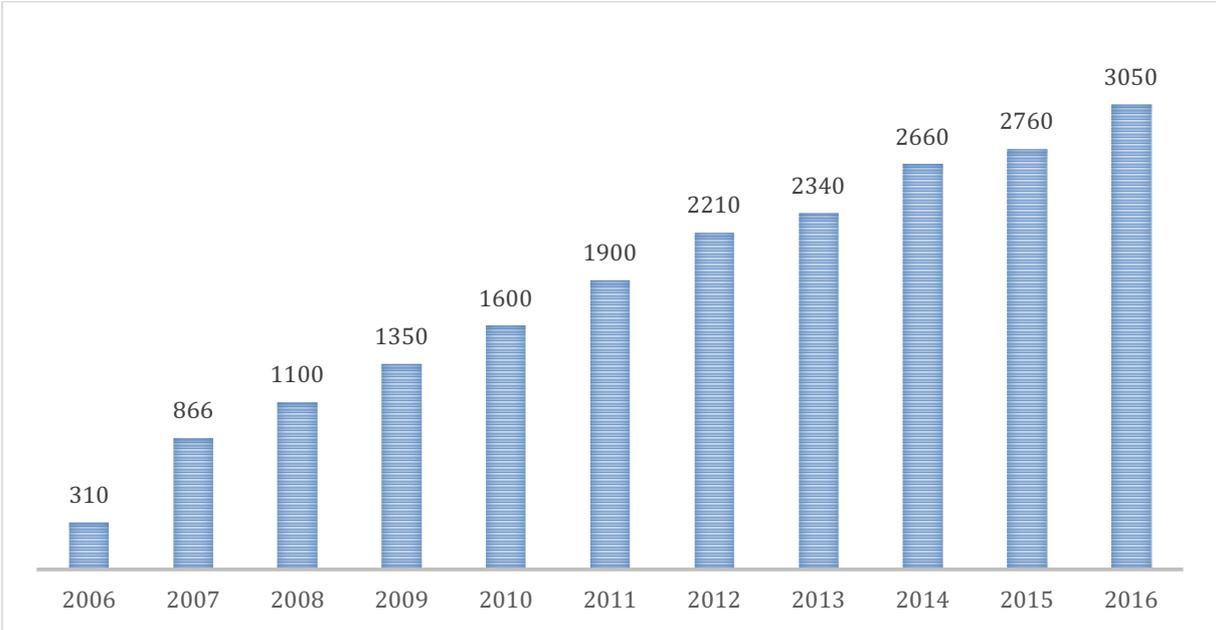

*Figure 3: Growth of the OpenDOAR meta-catalog database from growth worldwide OpenDOAR website (Feb 2017)*

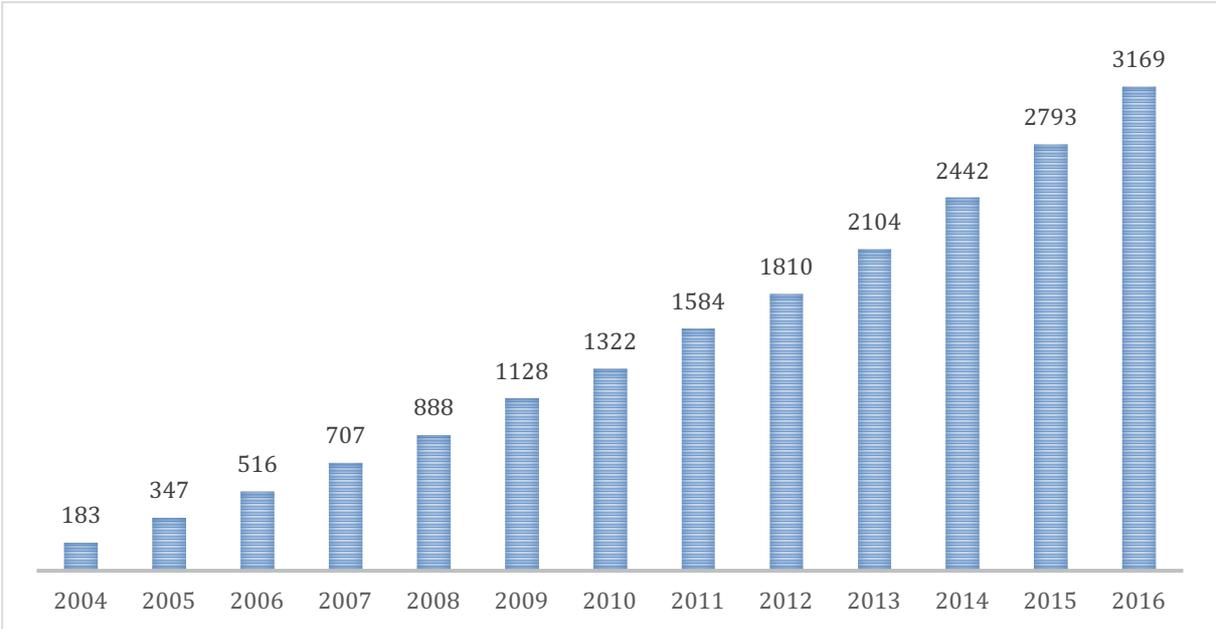

*Figure 4 : Growth of OpenArchives meta-catalog from Wayback Machine captures (February 2017)*

## Description of main repositories

*Table 2: Description of the meta-catalogs from official websites*

| Name and access | Date of creation | Coverage (Feb. 2017) | Geographical coverage | Search granularity | Selection modes and particularities |
|---|---|---|---|---|---|
| **OAIster** (OCLC, Michigan University) http://www.oclc.org/en/oaister.html | 2002 | 2000 catalogs | World | Simple and advanced search, no full text search (WorldCat interface) | - Submission by the institutions<br>- Contributors use the WorldCat Digital Collection to identify their repositories and synchronize their metadata with the WorldCat subset known as OAIster |
| **OpenArchives** (Digital Library Federation, Coalition for Networked Information, National Science Foundation Grant) https://www.openarchives.org | 2001 | 2799 catalogs | World | No search interface, only data providers listings | - Submission by the institutions |
| **OpenDOAR** (Open Society Institute, Joint | 2006 | 3315 catalogs | World | Google custom search | - Submission by the institutions |

| Name and access | Date of creation | Coverage (Feb. 2017) | Geographical coverage | Search granularity | Selection modes and particularities |
|---|---|---|---|---|---|
| Information Systems Committee, Consortium of Research Libraries, SPARCEurope) http://www.opendoar.org | | | | | - All the data providers have been visited by OpenDOAR staff<br>- Only data providers containing open-access resources (full text) are recorded (no catalog records) |
| **ROAR** (Joint Information Systems Committee) http://roar.eprints.org | 2003 | 4368 catalogs | World | - Google custom search<br>- Search by archive software, year, country and type of data provider | - Submission by the institutions<br>- Quality control by an « Editorial Review » |
| **OpenAIRE** (European Commission) https://www.openaire.eu | 2009 | 798 catalogs | Europe[3] | Ad hoc interface : faceted search | - Data providers must first be registered in OpenDOAR. |
| **University of Illinois (Illinois)** http://quest.library.illinois.edu/re | 2006 | 4415 catalogs | World | Search by word in the data provider name | - Data providers collected by the University of Illinois<br>- Submission by the institutions |

---

[3] N.B. : Some of the repositories on this platform can nevertheless be located outside the European territory because of the author's belonging to an academic institution elsewhere in the world.

| Name and access | Date of creation | Coverage (Feb. 2017) | Geographical coverage | Search granularity | Selection modes and particularities |
|---|---|---|---|---|---|
| gistry/searchform.asp | | | | | |

**Complete repository comparison**

For the most part, meta-catalogs were created during the first decade of the 2000s heeding to the multiplication of repositories all over the globe. They have systematically tried to organize and centralize these initiatives.
Table 2 proposes a synthesis of the six meta-catalogs' salient features that are perhaps considered as the most important ones, notably: OAIster, OpenArchives, OpenDOAR, ROAR, OpenAIRE and the University of Illinois OAI-PMH Data Provider Registry (Illinois). It should be noted that these meta-catalogs are not accessible through the OAI-PMH protocol nor are the selected repositories.

The worldwide coverage of meta-catalogs (except OpenAIRE) and their relative homogeneity announces possible redundancies in the number of repositories contained in the various meta-catalogs (approx. 2949 repositories per platform) with the exception of OAIster (2000 repositories) and OpenAIRE (798 repositories).

The content available within these various meta-catalogs cannot be clearly identified with the use of literature. Even though information on this subject is available on the respective sites of each meta-repository, it is fragmented or simply non-existent. (Confederation of open access repositories 2012) The administrators of the platforms that were contacted were sometimes able to provide this information, while at the same time conceding on several occasions the unfeasibility of precisely mapping the content. Nonetheless, a majority of (inter) institutional and multidisciplinary repositories (ROAR and OpenAIRE) can be observed; governmental and public repositories are also among the available resources, but in a negligible way (OpenDOAR).

**Search granularity and particularities**

The search tools contained within the platforms that are used for accessing the repository listings vary widely in granularity. They range from a simple search of the directory title (Illinois), to a much more refined faceted search (OpenAIRE), or even to an absence of a tool (OAI repositories). In some meta-catalogs (OpenDOAR and ROAR), there is a presence of the Google custom search tool enabling the direct access to documents included within the catalog. The relative insufficiency of the search tools emphasizes the difficulty of accessing the content held by meta-catalogs. Faced with the efficiency of tools like Google Scholar (Pedersen, Arendt 2014), it is clear that the search interfaces remain very insufficient (Norris, Oppenheim, Rowland 2008).

It appears that OpenDOAR occupies a designating role of reference within the range of studied meta-catalogs. Evidently, the repositories selected by the platform are consulted by OpenDOAR staff. This unique selection process (whose major criterion is the availability of open access resources[4]) between the meta-catalogs guarantees the activity of repositories, a certain level of quality and breaks with an automated approach. OpenAIRE is excluded here due to the requirements that its repositories be registered in OpenDOAR (Pinfield et al. 2014, p. 6-9).

---

[4] The list of criteria for inclusion and exclusion is available at :
http://www.opendoar.org/about.html#scope

**Harvesting of meta-catalogs and comparison**

After harvesting of the catalog of meta-catalogs, table 2 gives a first overview of the number of repositories available in each meta-catalog. After the headers, the first row "All items" indicates the total number of items found in each meta-catalog independently of the OAI-PMH selection (according to the used source, table 1). Then the second row indicates a selection of exclusively OAI-PMH catalogs. For instance, there is a strong difference for OpenDOAR between the first value of 3291 and the second value of 2249, which indicates that all repositories are not necessarily OAI-PMH servers but could also be RSS feeds (68.3% of OAI-PMH repositories in this study while Pinfield indicated 71% in 2014). According to our selection algorithm, a part of ROAR's repositories (80%) are also solely OAI-PMH (table 1, row 2). The third line of the table indicates the number of unique OAI-PMH repositories after the simple normalization and deduplication step described in the method section. In this case, only a few URLs were deduplicated, mainly ROAR (with 8%), as well as OAIster and OpenAIRE (but with less than 4%).

*Table 2: OAI-PMH repositories collected for each meta-catalog (January 2017)*

|           | OpenDOAR | ROAR | OpenArchives | Illinois | OAIster | OpenAIRE |
|-----------|----------|------|--------------|----------|---------|----------|
| All items | 3291     | 4365 | 2751         | **4659** | 1975    | -        |
| Only OAI  | 2249     | 3487 | 2751         | 4659     | 1975    | 764      |
| Unique    | 2235     | 3193 | 2751         | 4642     | 1912    | 745      |

In figure 5, the total number of OAI-PMH repositories is indicated under the ordinate in decreasing order (same numbers as indicated in the row 'Unique' of table 2). The Illinois meta-catalog is the largest one with 4642 links to OAI-PMH repositories, followed by ROAR with 3193. Furthermore, this figure indicates the active part of OAI-PMH server in blue and those that are not reachable using a simple OAI-PMH query ("?verb=Identify") in red. Due to its European perimeter of interest OpenAIRE is by far the smallest meta-catalog.

*Figure 5: reachable in blue and non-reachable in red numbers of repositories per meta-catalog (January 2017)*

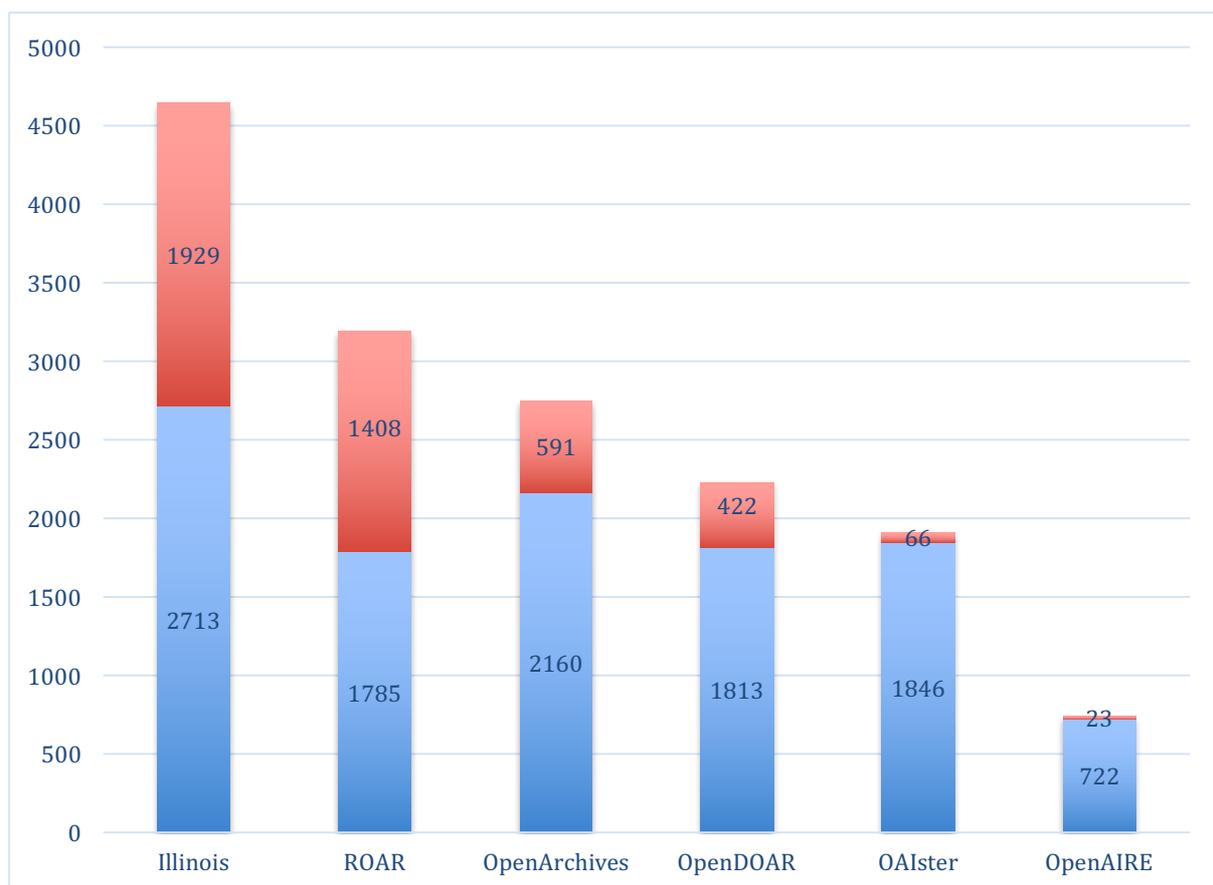

Table 3 provides detailed percentages of error and success per repository. The meta-catalogs containing the best percentage of operational OAI-PMH repositories are OpenAIRE and OAIster and the worst being ROAR with 55.9% (followed by Illinois 58.4%). Possessing a great concern for quality ROAR has a relatively good result of 81.1% of reachable OAI-PMH servers. The third line indicates the number of unique identifiers after deduplication (normalization of URLs). It can be observed that there are very few duplicate URLs. The one with the most duplicates was ROAR with 4% of the URL normalized. However, the normalization is also useful for the comparison of the URLs between meta-catalogs, which is the next important step.

*Table 3: Reachable OAI-PMH repositories among each meta-catalogs (January 2017)*

|  | OpenDOAR | ROAR | OpenArchives | Illinois | OAIster | OpenAIRE |
|---|---|---|---|---|---|---|
| Nb Total | 2235 | 3193 | 2751 | 4642 | 1912 | 745 |
| Nb Success | 1813 | 1785 | 2160 | 2713 | 1846 | 722 |
| Nb Unique | 1809 | 1722 | 2149 | 2666 | 1843 | 721 |
| % Success | 81.1 | **55.9** | 78.5 | 58.4 | **96.5** | **96.9** |
| % Error | 18.9 | 44.1 | 21.5 | 41.6 | 3.5 | 3.1 |

Table 4 shows the distribution of errors per http status code, however only errors > 1% appear. Code 500 (Server error) is the most frequent with 50% of occurrences, which means that in most cases, there are currently no active OAI-PMH repositories. The code 404 (Not Found), appears 26% of the time indicating that the OAI-PMH repository disappeared but their web server could still exists. Finally, in 18% of the case, some of the URLs are redirected to a web page and the code 200 (success) is returned, as the OAI-PMH repository has disappeared, these results are ultimately considered as errors in our study.

*Table 4: Percentage of error per http status code (January 2017)*

| http status code | Bad Request 400 | Forbidden 403 | Not Found 404 | Internal Server Error 500 | Service Unavailable 503 | Wrong success 200 |
|---|---|---|---|---|---|---|
| % of error per code | 0.02 | 0.01 | **0.26** | **0.5** | 0.02 | **0.18** |

**After cleaning and normalization**

A total of 4776 distinct OAI-PMH repository URLs were found amongst the six meta-catalogs.

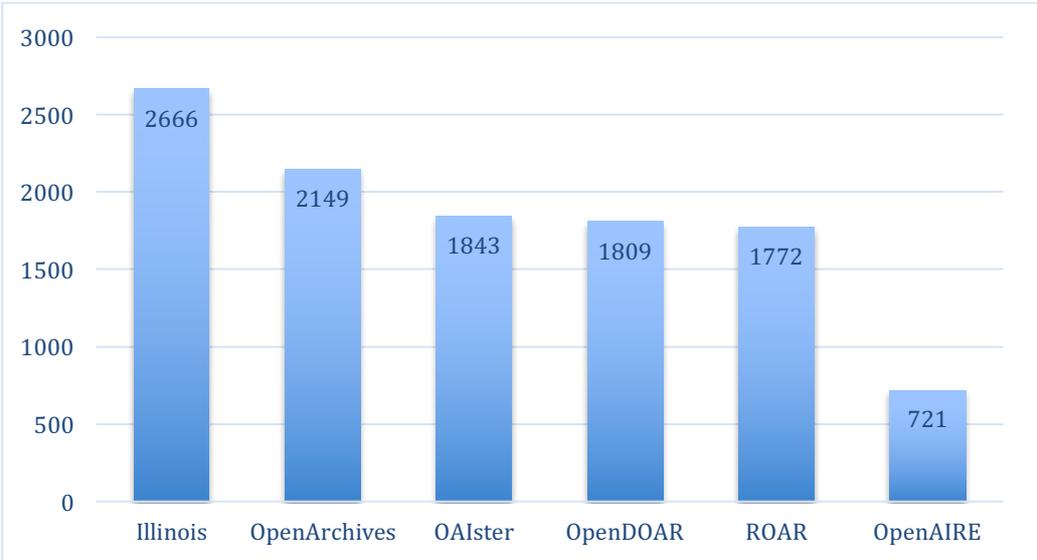

*Figure 6: Number of archives per repository after cleaning (January 2017)*

Figure 6 provides the number of operational OAI-PMH per meta-catalog after the normalization of URLs. In terms of the amount of OAI-PMH repositories, Illinois is still the top with 2666 active servers and OpenArchives comes in second. OAIster, OpenDOAR and ROAR are respectively close between them in terms of numbers. Due to its European perimeter OpenAIRE still ranks the lowest.

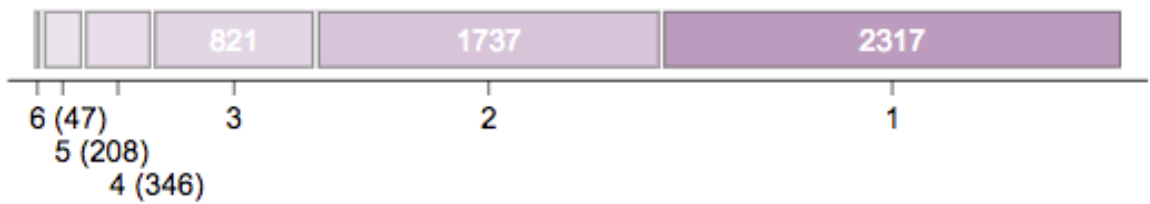

*Figure 7: Distribution of common repositories between all meta-catalogs (created with jvenn, Bardou 2014)*

Figure 7 shows which OAI-PMH repositories are common amongst the meta-catalogs. This figure is significant because it indicates that most of time OAI-PMH repositories are specific to a meta-catalog (2317 meaning 42.3%). And finally, 47 OAI-PMH repositories appear in the list of the 6 meta-catalogs (meaning less than 1%).

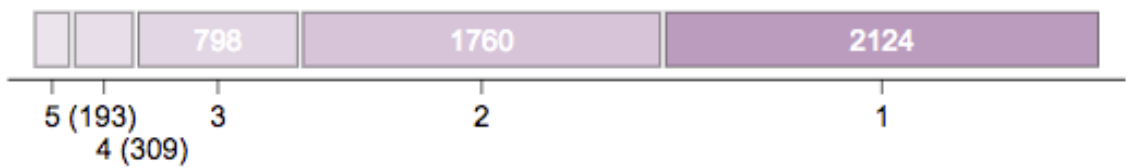

*Figure 8: Distribution of common repositories without OpenAIRE meta-catalog (created with jvenn, Bardou 2014)*

If the OpenAIRE meta-catalog (the European meta-catalog) is not considered and only the other five meta-catalogs are compared (figure 8), then it can be noted that there are still 40.9% of the repositories that remain specific to one meta-catalog.

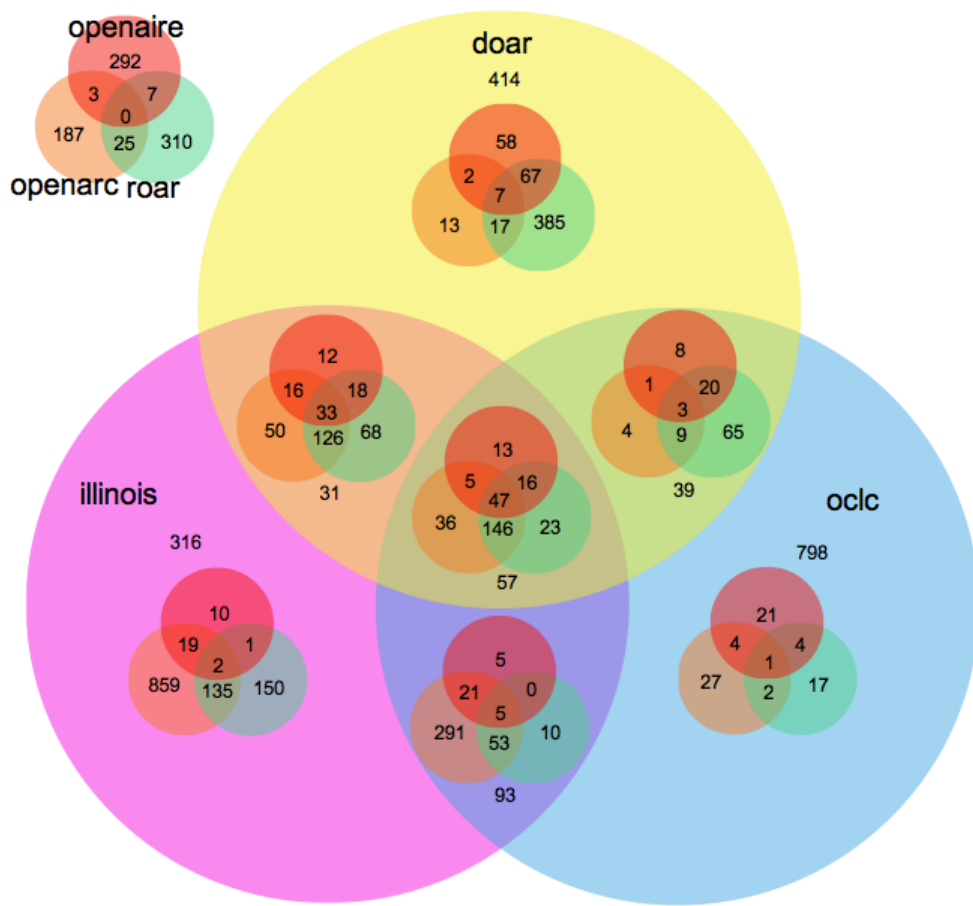

*Figure 9: Full Venn diagram between repositories, created with VainPainter, Lin 2016 (here, "oclc" is "OAIster" and "openarc" is "OpenArchive")*

Despite being challenging to interpret, figure 9 provides an exhaustive view of overlapping between the meta-catalogs. For instance, the number 47 can be observed at the center of the figure which is the number of OAI-PMH repositories common between all the studied meta-catalogs (results in table 7, in annex). Another interesting observation is the first number under OpenDOAR of 414 (7.5%), which is the number of OAI-PMH repositories solely specific to OpenDOAR (never appearing in other meta-catalogs). In the same scenario, for Illinois is 316 (5.8%), OAIster is 798 (14.6%), OpenAIRE is 292 (5.3%), OpenArchives is 187 (3.4%) and ROAR is 310 (5.6%).

Table 5 shows the comparison of overlapped numbers amongst each meta-catalog. This table is symmetrical except the last column that contains the total number OAI-PMH repositories of each line. OpenArchives and Illinois have the most elements in common with 1844. Afterwards, it was found that OpenDOAR and ROAR shared 1050 elements in common. Regarding OAIster, Illinois is the one with the most common elements. OpenAIRE is very specific but it can be observed that OpenDOAR has the most elements in common. Furthermore, it can surprisingly be noted that OAIster is the one with the least amount of elements in common with OpenDOAR and ROAR.

*Table 5: Matrice of common repositories between meta-catalog 2 on 2 (with total per line)*

|          | OpenDOAR | Illinois | OAIster | OpenAIRE | OpenArchives | ROAR | Total |
|---|---|---|---|---|---|---|---|
| OpenDOAR | 0 | 697 | **492** | 326 | 515 | **1050** | 1809 |
16

| | | | | | | | |
|---|---|---|---|---|---|---|---|
| Illinois | 697 | 0 | 821 | 223 | **1844** | 833 | 2666 |
| OAIster | 492 | **821** | 0 | 174 | 655 | **421** | 1843 |
| OpenAIRE | **326** | 223 | 174 | 0 | **169** | 231 | 721 |
| OpenArchives | 515 | **1844** | 655 | 169 | 0 | 611 | 2149 |
| ROAR | **1050** | 833 | 421 | 231 | 611 | 0 | 1772 |

Table 6 shows the comparison of overlapping ratio between meta-catalogs. Unlike table 4, this table is not symmetrical as it presents the ratio of common OAI-PMH repositories of two meta-catalogs compared to the total of OAI-PMH repositories of the first meta-catalog. For instance, the common data between OpenDOAR and ROAR represents 58% of OpenDOAR data (first line, last colon of the table) and 59% of ROAR data. But the largest ratios of common data are undoubtedly Openarchives and Illinois representing 86% and 69% of respectively. Contrarily, without taking OpenAIRE into consideration, OAIster and ROAR are the ones with the least in common, with 24% and 23% respectively (logically OAIster and OpenDOAR are in relatively the same situation). If OpenAIRE is once again taken into consideration, the ratio of common data with OpenDOAR is 45%.

*Table 6: Matrice of ratio of common repositories between meta-catalog 2 on 2 (based on total)*

| | OpenDOAR | Illinois | OAIster | OpenAIRE | OpenArchives | ROAR |
|---|---|---|---|---|---|---|
| OpenDOAR | 0 | 0.39 | **0.27** | 0.18 | **0.28** | **0.58** |
| Illinois | **0.26** | 0 | 0.31 | 0.08 | **0.69** | 0.31 |
| OAIster | 0.27 | **0.45** | 0 | 0.09 | 0.36 | **0.23** |
| OpenAIRE | **0.45** | 0.31 | **0.24** | 0 | **0.23** | 0.32 |
| OpenArchives | **0.24** | **0.86** | 0.3 | 0.08 | 0 | 0.28 |
| ROAR | **0.59** | 0.47 | **0.24** | 0.13 | 0.34 | 0 |

# Limitations

**Surface analysis**
For the purpose of this article, only the OAI-PMH repository level was studied. Thus, there was no concern surrounding the number of records in the OAI-PMH repositories or the content itself. However, it can certainly be assumed that there should be a Gaussian distribution in OpenDOAR repositories (Pinfield 2014). Furthermore, the authors added: "There is also a low number of very small repositories, with less than 6.6% of the population having 100 or fewer items. This is not surprising given that initial repository deployment may often tend to be associated with deposits of an early tranche of materials and organizations may not invest in initiating repository deployment without such a corpus being available" (2014, p. 2406). With 42.3% specificity, it would be interesting to know if there is a type of content specific to a meta-catalog which could come from policy, method, or operational way of inclusion. Nevertheless, this is out of the scope of this study but could be a future research case.

**Methods for observing OAI-PMH evolution**
One of the original concerns relates to the viability of the OAI-PMH infrastructure, or in other words whether OAI-PMH is still active and used by the librarian community. This study represents just one facet of this question; one efficient method could be to study the real usage in terms of queries made to the main OAI-PMH repositories throughout time. Consequently, Archived Logs of the server could be a valuable source for this type of analysis.

**Availability of the OAI-PMH services**
Regarding the access to the OAI-PMH server itself, when possible the harvester that was used was improved aiding its recognition. Firstly, a user-agent name was added, and secondly a replicated web browser name was sometime needed. Furthermore, the crawler was modified to allow it to also accept https connections. In the end, some services were temporarily unavailable and the evaluation had to be done over a 3-day period.

# Discussions

This study clearly shows that OAI-PMH repositories continue to grow despite a decline of interest from the research world. Among all collected OAI-PMH repositories, 50% of them are not operational but there are significant differences between the considered meta-catalogs. Such as OAIster and OpenAIRE containing the best up to date data and OpenDOAR and OpenArchives also having good results. Quantitatively, Illinois and OpenArchives are the best with more than 2000 entries, although OpenArchives surpasses this good result by having less errors compared to ROAR or Illinois meta-catalogs. Currently there is a total 4776 distinct operational OAI-PMH repositories listed in these six meta-catalogs. Surprisingly, all the studied meta-catalogs contain 42.3% specific OAI-PMH repositories (even without considering OpenAIRE) and less than 1% is that of common data amongst them. The meta-catalogs with the most entries in common are Illinois and OpenArchives (between 86% and 69%). But also, ROAR and OpenDOAR have a significant list of entries in common (between 58% and 59%). However, OAIster is quite isolated from the other meta-catalogs (except Illinois in one direction).

This study was challenging to conduct due to the heterogeneity of access to the meta-catalogs. A standard protocol or service to acquire different repository lists would have been desirable. One way to achieve this could be to directly present entries through the OAI-PMH services. Due to high a specificity rate of 42.3%, another strong recommendation would be to have a generally updated meta-catalog that could contain all the repositories appearing in the different studied meta-catalogs. Of course, one of the current catalogs could be entrusted with this mission.

Ultimately, the last step for the end user is to provide a state of the art search engine containing all the documents listed in the repositories (even a web service for the search). This state of art search engine could be defined as a tool that works as simply as Google/Google Scholar with full text (when available), while also being able to support meta-data such as the WOS search engine (also fast and relevant). Currently three out of six of meta-catalogs provide search engines of their repositories content: 1/ OpenDOAR, through a Google Custom Search Engine (since 2006, smart, fast, full-text, but no meta-data and risk of indexing gap), 2/OAIster, with WorldCat services on OAIster repositories only data (well-designed but no full-text[5]) and, 3/ OpenAIRE search engine (European only, well-designed but no full-text). Furthermore, all three of these search engines suffer from a lack of exhaustively (identified here in this article). Furthermore, McCown (2006) and Perdersen (2014), give a coverage for academic production of 44% for Google and 55% for Google Scholar respectively (in computer sciences). According to Norris' findings in 2008, Google Scholar was better for open access production coverage with 68% for OAIster compared to 9.62% for OpenDOAR. Considering all the previously discussed points, there is undoubtedly a place for a state of art search engine which could have better coverage, transparency, full text search and meta-data structuration.

## Conclusions

This is the first study that attempts to evaluate the OAI-PMH eco-system on such a large scale. Ultimately, it can be observed that the OAI-PMH eco-system continues to progressively disseminate documents, notably scholarly and open access documents. Although, two out of the six meta-catalogs list of entries should be cleaned. There is a very high rate of cumulated specificity (42%) which indicates the necessity to propose a general meta-catalog able to contain it all. Furthermore, a more exhaustive and transparent search engine would be beneficial for end users and the librarian community, especially when considering the gaps of Google Scholar and meta-catalogs independently.

---

[5] Empirically tested with content queries of indexed article

Annex:

Table 2: List of 47 omnipotent OAI-PMH repositories (present in the six repositories)

| |
|---|
| bdigital.unal.edu.co/cgi/oai2 |
| biecoll.ub.uni-bielefeld.de/oai2/oai2.php |
| bvh.univ-tours.fr/oai2/repositoryOAI.asp |
| digital.ub.uni-paderborn.de/oai |
| docserv.uni-duesseldorf.de/servlets/OAIDataProvider |
| documentation.ird.fr/fdi/oai.php |
| drops.dagstuhl.de/opus/phpoai/oai2.php |
| dspace.nwu.ac.za/oai/request |
| earchive.tpu.ru/oai/request |
| econstor.eu/dspace-oai/request |
| edoc.hu-berlin.de/OAI-2.0 |
| elar.rsvpu.ru/oai/request |
| elar.urfu.ru/oai/request |
| elar.usfeu.ru/oai/request |
| elar.uspu.ru/oai/request |
| elib.uraic.ru/oai/request |
| epic.awi.de/cgi/oai2 |
| eprints-phd.biblio.unitn.it/cgi/oai2 |
| eprints.ucm.es/cgi/oai2 |
| eprints.unife.it/cgi/oai2 |
| eprints.uniss.it/cgi/oai2 |
| epub.wu.ac.at/cgi/oai2 |
| funes.uniandes.edu.co/cgi/oai2 |
| gala.gre.ac.uk/cgi/oai2 |
| helvia.uco.es/oai/request |
| jultika.oulu.fi/OAI/Server |
| kluedo.ub.uni-kl.de/oai |
| libros.metabiblioteca.org/oai/request |
| lup.lub.lu.se/oai |
| monarch.qucosa.de/oai |
| oai.bibliothek.uni-kassel.de/dspace-oai/request |
| oceanrep.geomar.de/cgi/oai2 |
| openaccess.city.ac.uk/cgi/oai2 |
| orbi.ulg.ac.be/oai/request |
| orbilu.uni.lu/oai/request |
| paduaresearch.cab.unipd.it/cgi/oai2 |
| pedocs.de/oai2/oai2.php |
| pure.qub.ac.uk/ws/oai |
| qspace.qu.edu.qa/oai/request |
| qucosa.de/oai |
| repository.eafit.edu.co/oai/request |
| serval.unil.ch/oaiprovider |
| ssoar.info/OAIHandler/request |
| tesionline.unicatt.it/dspace-oai/request |
| tuprints.ulb.tu-darmstadt.de/cgi/oai2 |
| veprints.unica.it/cgi/oai2 |
| zora.uzh.ch/cgi/oai2 |